\documentclass[preprint,authoryear,12pt]{elsarticle}
\usepackage{times}
\usepackage{amsmath, amsthm, amssymb, amstext, amsfonts}
\usepackage{appendix}
\usepackage{array}
\usepackage{bm}
\usepackage{booktabs,threeparttable}
\usepackage{color}
\usepackage{comment}
\usepackage{courier}
\usepackage{dcolumn}
\usepackage{enumerate}
\usepackage{fancyhdr}
\usepackage{fancyvrb}
\usepackage{float}
\usepackage[margin=3cm]{geometry}
\usepackage{graphicx}
\usepackage[bookmarks=False, hidelinks, pdfstartview={XYZ null null 1.00}]{hyperref}
\usepackage{ifpdf}
\usepackage{listings}
\usepackage{mathrsfs}
\usepackage{moreverb}
\usepackage{multirow}
\usepackage{natbib}
\usepackage{placeins}
\usepackage{slashbox}
\usepackage{stackrel}
\usepackage[caption=false]{subfig}
\usepackage{tabularx}
\usepackage{times}
\usepackage{xcolor}
\usepackage{bbm}

\newtheoremstyle{remboldstyle}
  {}{}{\itshape}{}{\bfseries}{.}{.5em}{{\thmname{#1 }}{\thmnumber{#2}}{\thmnote{ (#3)}}}
\theoremstyle{remboldstyle}
\theoremstyle{definition}

\theoremstyle{definition}

\let\oldproofname=\proofname
\renewcommand{\proofname}{\rm\bf{\oldproofname}}

\makeatletter
\def\Eqlfill@{\arrowfill@\Relbar\Relbar\Relbar}
\newcommand{\extendEql}[1][]{\ext@arrow 0359\Eqlfill@{#1}}
\makeatother

\makeatletter

\newcommand{\Rmnum}[1]{\expandafter\@slowromancap\romannumeral #1@}
\makeatother

\newcommand{\spc}{\hspace{2 mm}}

\newcommand{\ofx}{\left(\boldsymbol{x}\right)}
\newcommand{\di}{\mathrm{d}}

\newcommand{\bx}{\boldsymbol{x}}
\newcommand{\by}{\boldsymbol{y}}
\newcommand{\br}{\boldsymbol{\rho}}
\newcommand{\bgamma}{\boldsymbol{\gamma}}
\newcommand{\argmin}{\operatornamewithlimits{argmin}}
\newcommand{\argmax}{\operatornamewithlimits{argmax}}

\newcolumntype{C}[1]{>{\centering\let\newline\\\arraybackslash\hspace{0pt}}m{#1}}

\newcolumntype{.}{D{.}{.}{-1}}

\newlength{\struthd}

\bibpunct{(}{)}{;}{a}{}{,}

\begin{document}
\graphicspath{{Figures//}}
\begin{frontmatter}

\title{Variable Selection in Bayesian Semiparametric Regression Models}

\author[TAU]{Ofir Harari}
\ead{ofirhara@post.tau.ac.il}
\author[TAU]{David M. Steinberg}
\ead{dms@post.tau.ac.il}
\address[TAU]{Department of Statistics and Operations Research,
Raymond and Beverly Sackler Faculty of Exact Sciences, Tel Aviv University, Tel Aviv 69978, Israel}
\begin{abstract}
In this paper we extend existing Bayesian methods for variable
selection in Gaussian process regression, to select both the
regression terms and the active covariates in the spatial
correlation structure. We then use the estimated posterior
probabilities to choose between relatively few modes through cross-validation, 
and consequently improve prediction.
\end{abstract}
\end{frontmatter}

\section{Introduction}
\noindent Gaussian processes have been employed extensively both in
regression and classification problems in machine learning (see
\citealt{Rasmussen}). Their earliest use in statistical modeling may
have been for fitting Kriging predictors in geostatistics (see e.g. \citealt{Krige1951}, \citealt{Cressie})
and they have been the most popular metamodeling approach to computer simulations (see \citealt[chapter~3]{santner}).\\

\noindent The problem of variable selection in Gaussian process regression has been addressed by \cite{Bingham}, while \cite{Vannucci}
extended their results to generalized linear nonparametric models. In both cases no regression terms were included, and dependence on the factors was modeled by the authors solely
via the spatial correlation structure.\\

\noindent In many applications only interpolation or smoothing are required,
and the omission of regression terms can simplify the fitting of the Kriging metamodel,
with little expense in predictive precision (see e.g. \citealt{IMSE}).  It is also known that these models have
a strong ``Kriging towards the mean'' tendency at sites far away from the training data, and it is thus
often essential to add a trend for effective extrapolation (see \citealt{trend} for the original argument). Currently,
the common practice is to either include all of the covariates both in the correlation structure and the linear component (i.e. {\it Universal Kriging}) or to omit the linear part altogether and model the data as observations from a pure (weakly) stationary random process (i.e. {\it Ordinary Kriging}). \cite{BlindKrig} proposed a Bayesian procedure for designed computer experiments, to select first and second order polynomials for the mean surface while leaving the covariance terms untouched. In this work we allow for exclusion of covariates from the correlation structure, along with their possible inclusion in the trend component, in the case that their only contribution to the output is linear.

\section{Model Assumptions}
\begin{figure}[ht]
  \centering
 \includegraphics[width=.5\linewidth]{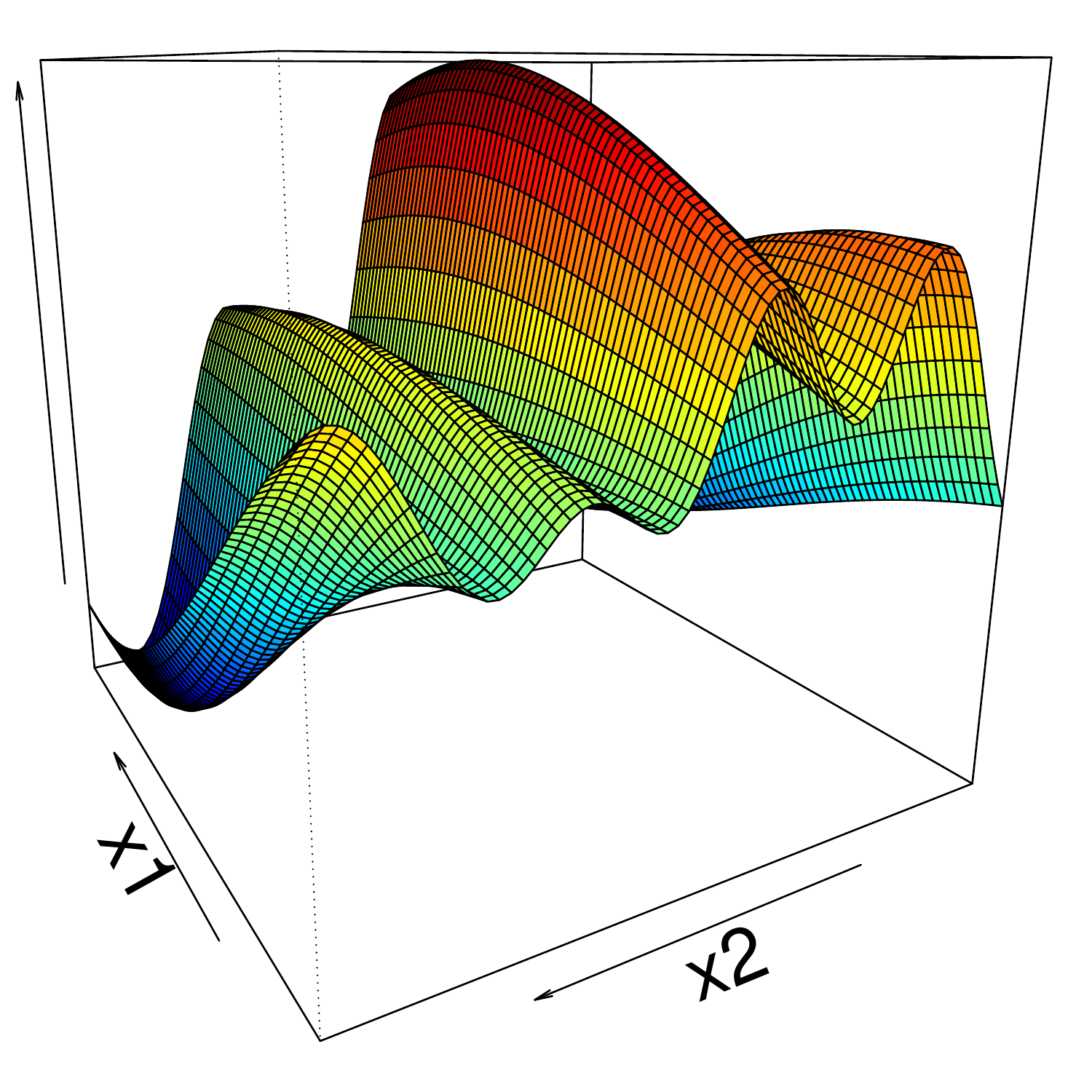}
  \caption{A realization of $Z\ofx$ with the correlation function (\ref{eq:corr_func}), for $p=2$,$\rho_1= 0.75$ and $\rho_2=5\times10^{-6}$.}
  \label{fig:3D_Realization}
\end{figure}

\noindent Let $\left\{\boldsymbol{x}_1,\ldots,\boldsymbol{x}_n\right\}\subset\mathcal{X}$ be our training set, where $\mathcal{X}\subset\mathbb{R}^p$ is the experimental region, and let $\boldsymbol{y} = \left[y\left(\boldsymbol{x}_1\right),\ldots,y\left(\boldsymbol{x}_n\right)\right]^{\mathsf{T}}$ be a vector of observations from the model
\begin{align}\label{eq:model}
y\ofx = \beta_0 + \bx^{\mathsf{T}}\boldsymbol{\beta} + Z\ofx + \varepsilon\ofx\spc,
\end{align}
where $\beta_0$ is a constant, $\boldsymbol{\beta} = \left[\beta_1,\ldots,\beta_p\right]^{\mathsf{T}}$ is a vector of regression coefficients, $Z\ofx$ is a zero mean stationary Gaussian process with marginal variance $\sigma^2_{_Z}$ and a separable correlation function of the form
\begin{align}\label{eq:corr_func}
R\left(\boldsymbol{u},\boldsymbol{v};\boldsymbol{\rho}\right) = \prod_{j=1}^p\rho_j^{\left(u_j-v_j\right)^2}
\end{align}
for some vector of correlation parameters $\boldsymbol{\rho} = \left[\rho_1,\ldots,\rho_p\right]^{\mathsf{T}}$, $0\leq\rho_j\leq1$ , $1\leq j\leq p$, and $\varepsilon\ofx$ is a white noise process, independent of $Z\ofx$, with variance parameter $\sigma^2_{\varepsilon}$. Note that the closer $\rho_j$ is to $1$, the less $Z\ofx$ tends to vary in the $j$th direction, and vice versa. This point is illustrated in Figure \ref{fig:3D_Realization}, in which a single realization of $Z\ofx$ is drawn, with $p=2$, a relatively large value of $\rho_1$ and an extremely small value of $\rho_2$.\\

\noindent Denoting $\lambda = \sigma^2_{\varepsilon}/\sigma^2_{_Z}$, it can be easily verified that
\begin{align}\label{eq:likelihood}
\by\big|\beta_0,\boldsymbol{\beta},\sigma^2_{_Z},\br,\lambda\sim\mathcal{N}\left\{\beta_0\boldsymbol{1}_n + \boldsymbol{X}\boldsymbol{\beta}\hspace{.1em},\hspace{.1em}\sigma^2_{_Z}\left[\boldsymbol{R}\left(\br\right)+\lambda\boldsymbol{I}\right]\right\}\spc,
\end{align}
where $\boldsymbol{X}$ is the design matrix and $$\boldsymbol{R}_{lm}\left(\br\right) = R\left(\bx_l,\bx_m\right) = \displaystyle\prod_{j=1}^p\rho_j^{(\bx_{lj}-\bx_{mj})^2}\spc.$$

\noindent We can interpret (\ref{eq:model}) as follows: each of the $p$ covariates in $\bx$ may (or may not) make a linear contribution to the response $y\ofx$. If, in addition, it makes a nonlinear contribution, it should be captured by $Z\ofx$, whose flexible nature makes it free of any rigid structure. The white noise term, also known as the ``nugget'' (see \citealt{nugget}) and interpreted as ``measurement error'', is often omitted when using the Kriging metamodel to emulate deterministic computer simulations. However, its (possibly unnatural) inclusion has certain benefits, both numerical and in terms of the average squared prediction error. The consequent $\lambda$ term in (\ref{eq:likelihood}) is mostly
identified with the equivalent penalized least squares minimization problem in the Reproducing Kernel Hilbert Space corresponding to $R(\cdot,\cdot)$ (see \citealt{Wahba}). In this paper, we would like to examine the possibility of leaving some of the covariates out of the correlation function. By doing that, we deliberately omit variables that are (in the case of computer experiments) known for a fact to be part of the computer code. It is then reasonable to add an error term, to avoid interpolation at the learning set, in favor of better predictions at new sites.

\section{Formalizing as a Bayesian Model}\label{sec:Bayesian}
\noindent We would now like to turn our attention to the problem of variable selection in model (\ref{eq:model}). Research on Bayesian variable selection in linear regression models has been thoroughly documented - \cite{Chipman} give a detailed account of some theoretical and computational aspects 
\cite{Bingham} and \cite{Vannucci} apply point-mass mixture prior distributions for the correlation parameters of simple Kriging models.
In this work, we aim to propose a unified framework, in which ``spike-and-slab'' priors, like those used by \cite{Bingham} and \cite{Vannucci}, are employed to select the components of both the regression and spatial parts of the model.

\subsection{Writing the Posterior Distribution}
\noindent To put the above to practice, we now assume that the regression coefficients are independent, with prior distributions
\begin{align}\label{eq:beta_prior}
\beta_j\big|\omega_r \sim \omega_r\mathcal{N}\left(0,\tau^2\right) + \left(1-\omega_r\right)\delta_0\spc,
\end{align}
where $\delta_0$ is a point mass at zero. Alternatively, one can denote by $\bgamma^r=\left(\gamma^r_1,\ldots,\gamma^r_p\right)$ a vector of latent variables, in which $\gamma^r_j=\mathbbm{1}\left\{\beta_j\ne0\right\}$ indicates whether or not the $j$th covariate is included in the linear component. We may then write $\gamma^r_j\big|\omega_r\sim\mathrm{Binom}\left(1,\omega_r\right)$ and $\beta_j\big|\gamma^r_j\sim\mathcal{N}\left(0,\gamma^r_j\tau^2\right)$ .\\

\noindent Regarding the spatial component, for the $j$th variable to be inert, its corresponding correlation factor $\rho_j$ must be equal to $1$. An analogue of (\ref{eq:beta_prior}) would then be
\begin{align}\label{eq:rho_prior}
\rho_j\big|\omega_c\sim \omega_c\hspace{.1em}\mathcal{U}(0,1) + \left(1-\omega_c\right)\delta_1\spc,
\end{align}
implemented by defining the vector of latent variables $\bgamma^c=\left(\gamma^c_1,\ldots,\gamma^c_p\right)$, where $\gamma^c_j\big|\omega_c\sim\mathrm{Binom}\left(1,\omega_c\right)$ and $$\pi\left(\rho_j\big|\gamma^c_j\right) = \gamma^c_j\hspace{.1em}\mathcal{U}(0,1) + \left(1-\gamma^c_j\right)\delta_1\left(\rho_j\right)\spc.$$

\noindent Using Bayes rule, we may now write the posterior distribution
\begin{align}\label{eq:posterior}
p&\left(\bgamma^r,\bgamma^c,\omega_r,\omega_c,\beta_0,\boldsymbol{\beta},\br,\sigma^2_{_Z},\lambda\big|\by\right)\nonumber\\[1em]
 &\propto p\left(\by\big|\beta_0,\boldsymbol{\beta},\sigma^2_{_Z},\br,\lambda\right)\pi\left(\bgamma^r,\boldsymbol{\beta}\big|\omega_r\right)\pi\left(\bgamma^c,\br\big|\omega_c\right)
 \times\pi\left(\beta_0,\sigma^2_{_Z},\lambda,\omega_r,\omega_c\right)\spc,
\end{align}
where $p\left(\by\big|\beta_0,\boldsymbol{\beta},\sigma^2_{_Z},\br,\lambda\right)$ is the multivariate Normal density, as indicated in (\ref{eq:likelihood}),
\begin{align*}
\pi\left(\bgamma^r,\boldsymbol{\beta}\big|\omega_r\right)= \omega_r^{\sum_{j}\mathbbm{1}\left\{\gamma^r_j=1\right\}}\left(1-\omega_r\right)^{\sum_{j}\mathbbm{1}\left\{\gamma^r_j=0\right\}}\prod_{\gamma^r_j=1}\phi_{\tau}\left(\beta_j\right)
\end{align*}
(here $\phi_{\tau}\left(\beta_j\right)$ is the Normal pdf with zero mean and standard deviation $\tau$, evaluated at $\beta_j$) and
\begin{align*}
\pi\left(\bgamma^c,\boldsymbol{\rho}\big|\omega_c\right) = \omega_c^{\sum_{j}\mathbbm{1}\left\{\gamma^c_j=1\right\}}\left(1-\omega_c\right)^{\sum_{j}\mathbbm{1}\left\{\gamma^c_j=0\right\}}\spc.
\end{align*}

\subsection{Markov Chain Monte Carlo Sampling from the Posterior Distribution}\label{sec:MCMC}
\noindent We now apply a Metropolis Sampling algorithm to draw a random sample from the posterior distribution. A
suitable `jump' (or `proposal') distribution is required - one that will allow the resulting Markov chain to arrive at any
possible state with positive probability. Here we extend the key ideas of \cite{Vannucci}, to propose a full Metropolis-Hastings schema:
\begin{enumerate}
\item
Update $\left(\bgamma^r,\boldsymbol{\beta}\right)$ and $\left(\bgamma^c,\boldsymbol{\rho}\right)$ : draw $k\in\left\{0,1,\ldots,2p\right\}$
from a $\mathrm{Binom}(2p,\nu)$ distribution (typically $\nu=\left(2p\right)^{-1}$) and change the present state of $k$ randomly chosen latent variables out of
$\left(\bgamma^r,\bgamma^c\right)$.
\begin{itemize}
\item[-]
If the value of $\gamma_j^r$ changes from $0$ to $1$, sample $\tilde{\beta}_j$ from a $\mathcal{N}\left(0,\tau^2\right)$ proposal.
\item[-]
If the value of $\gamma_j^c$ changes from $0$ to $1$, sample $\tilde{\rho}_j$ from a $\mathcal{U}(0,1)$ proposal.
\item[-]
If the value of $\gamma_j^r$ changes from $1$ to $0$, set $\tilde{\beta}_j=0$. This results in covariate
$\bx_j$ being omitted from the linear trend in the current iteration.
\item[-]
If the value of $\gamma_j^c$ changes from $1$ to $0$, set $\tilde{\rho}_j=1$. This results in covariate
$\bx_j$ being omitted from the correlation function in the current iteration.
\item[*]
In any of the above cases, add a small jitter to the remaining $\beta$ and $\rho$ values (while making sure all the $\rho_j$-s remain in the interval $(0,1)$).
\end{itemize}
\item
Update $\left(\beta_0,\sigma^2_{_Z},\lambda,\omega_r,\omega_c\right)$ : to simplify matters, transform the parameters to $$\mu = \log\left(\sigma^2_{_Z}\right) \spc;\spc \zeta = \log(\lambda)\spc;\spc \psi_r = \log\frac{\omega_r}{1-\omega_r}$$ and $$\psi_c = \log\frac{\omega_c}{1-\omega_c}\spc,$$ and sample $(\tilde{\beta}_0,\tilde{\mu},\tilde{\zeta},\tilde{\psi}_r,\tilde{\psi}_c)$ from a multivariate Normal distribution about the present state, $(\beta_0,\mu,\zeta,\psi_r,\psi_c)$. In our implementation we used independent components. Remember to take the Jacobian term $$\dfrac{\partial\left(\sigma^2_{_Z},\lambda,\omega_r,\omega_c\right)}{\partial{\left(\mu,\zeta,\psi_r,\psi_c\right)}} = \mu\hspace{.1em}\zeta\hspace{.1em}\omega_r(1-\omega_r)\hspace{.1em}\omega_c(1-\omega_c)$$ into account in the posterior later.
\item
The proposed vector $\left(\tilde{\bgamma}^r,\tilde{\boldsymbol{\beta}},\tilde{\bgamma}^c,\tilde{\boldsymbol{\rho}},\tilde{\beta}_0,\tilde{\sigma}^2_{_Z},\tilde{\lambda},\tilde{\omega}_r,\tilde{\omega}_c\right)$ is accepted with probability
\begin{align}\label{eq:accept}
\alpha = \min\left\{1\hspace{.1em},\hspace{.1em}\dfrac{p\left(\tilde{\bgamma}^r,\tilde{\bgamma}^c,\tilde{\beta}_0,\tilde{\boldsymbol{\beta}},\tilde{\br},\tilde{\sigma}^2_{_Z},\tilde{\lambda},\tilde{\omega}_r,\tilde{\omega}_c\big|\by\right)}{p\left(\bgamma^r,\bgamma^c,\beta_0,\boldsymbol{\beta},\br,\sigma^2_{_Z},\lambda,\omega_r,\omega_c\big|\by\right)}\right\}\spc.
\end{align}
Note that (\ref{eq:accept}) holds, since by construction, the proposal distribution described above is symmetric, i.e.
\begin{align*}
p\left(\tilde{\bgamma}^r,\tilde{\boldsymbol{\beta}},\tilde{\bgamma}^c,\tilde{\br}\big|\bgamma^r,\boldsymbol{\beta},\bgamma^c,\br\right)=p\left(\bgamma^r,\boldsymbol{\beta},\bgamma^c,\br\big|\tilde{\bgamma}^r,\tilde{\boldsymbol{\beta}},\tilde{\bgamma}^c,\tilde{\br}\right)
\end{align*}
and
\begin{align*}
p\left(\tilde{\beta}_0,\tilde{\sigma}^2_{_Z},\tilde{\lambda}\big|\beta_0,\sigma^2_{_Z},\lambda\right)= p\left(\beta_0,\sigma^2_{_Z},\lambda,\omega_r,\omega_c\big|\tilde{\beta}_0,\tilde{\sigma}^2_{_Z},\tilde{\lambda},\tilde{\omega}_r,\tilde{\omega}_c\right)\spc.
\end{align*}
\end{enumerate}

\section{Prediction Strategies}\label{sec:prediction}
\noindent We now wish to take advantage of the procedure described in Section \ref{sec:MCMC}, to make accurate predictions at $m$ new sites $\boldsymbol{X}_{\text{new}}$ . We will propose two possible approaches, the first of which - based on model averaging -  is a more ``orthodox'' Bayesian approach, while the second one, in which we merely use the schema of Section \ref{sec:MCMC} to identify the MAP model, might perhaps be characterized as opportunistic.

\subsection{Prediction based on Model Averaging}
\noindent Conditional on $\boldsymbol{\Theta} = \left(\beta_0,\boldsymbol{\beta},\br,\sigma^2_{_Z},\lambda\right)$, the joint distribution of $\boldsymbol{y}_{\mathrm{old}} = y\left(\boldsymbol{X}_{\mathrm{old}}\right)$ and $\boldsymbol{y}_{\mathrm{new}} = y\left(\boldsymbol{X}_{\mathrm{new}}\right)$ is given by
$$
\left[
\begin{array}{c}
\boldsymbol{y}_{\mathrm{new}}\\
\hline
\boldsymbol{y}_{\mathrm{old}}
\end{array}
\right]=
\mathcal{N}\left\{\left[
\begin{array}{c}
\beta_0\boldsymbol{1}_m + \boldsymbol{X}_{\mathrm{new}}\boldsymbol{\beta}\\
\hline
\beta_0\boldsymbol{1}_n + \boldsymbol{X}_{\mathrm{old}}\boldsymbol{\beta}
\end{array}
\right]\spc;\spc
\boldsymbol{\Sigma}\right\}\spc,
$$
where
$$\boldsymbol{\Sigma}=
\sigma^2_Z
\left[
\begin{array}{c|c}
\boldsymbol{R}_{\mathrm{new}} + \lambda\boldsymbol{I}_m &\boldsymbol{R}_{\mathrm{new},\mathrm{old}}\\
\hline
\boldsymbol{R}_{\mathrm{new},\mathrm{old}}^{\mathsf{T}} &\boldsymbol{R}_{\mathrm{old}} + \lambda\boldsymbol{I}_n
\end{array}
\right]
$$
and
$\boldsymbol{R}_{\mathrm{new},\mathrm{old}}$ is the cross-correlation matrix $$\boldsymbol{R}_{\mathrm{new},\mathrm{old}}^{i,j} = R\left(\boldsymbol{x}_{\mathrm{new}}^i,\boldsymbol{x}_{\mathrm{old}}^j\right)\spc.$$

\noindent It is therefore straightforward to show that
\begin{align}\label{eq:pred_conditional}
\mathbb{E}\left\{\boldsymbol{y}_{\mathrm{new}}\Big|
\boldsymbol{y}_{\mathrm{old}},\boldsymbol{\Theta}\right\}&=
\beta_0\boldsymbol{1}_m+\boldsymbol{X}_{\mathrm{new}}\boldsymbol{\beta}\nonumber \\[1em]
&+\boldsymbol{R}_{\mathrm{new},\mathrm{old}}\left[\boldsymbol{R}_{\mathrm{old}} + \lambda\boldsymbol{I}_n\right]^{-1}\left(\boldsymbol{y}_{\mathrm{old}}-\beta_0\boldsymbol{1}_n - \boldsymbol{X}_{\mathrm{old}}\boldsymbol{\beta}\right)\spc,
\end{align}
and predictions, based on a sample $\left\{\boldsymbol{\Theta}_i\right\}_{i=1}^N$ from the posterior distribution (as described in section \ref{sec:MCMC}), can be made by integrating $\boldsymbol{\Theta}$ out, namely
\begin{align}\label{eq:pred_averaged}
\widehat{\boldsymbol{y}}_{\mathrm{new}} &=\mathbb{E}\left\{\boldsymbol{y}_{\mathrm{new}}\Big|
\boldsymbol{y}_{\mathrm{old}}\right\}= \int\mathbb{E}\left\{\boldsymbol{y}_{\mathrm{new}}\Big|
\boldsymbol{y}_{\mathrm{old}},\boldsymbol{\Theta}\right\}p\left(\boldsymbol{\Theta}\big|\boldsymbol{y}_{\mathrm{old}}\right)\di\boldsymbol{\Theta}\nonumber\\[1em]
&\approx \frac{1}{N}\sum_{i=1}^N\mathbb{E}\left\{\boldsymbol{y}_{\mathrm{new}}\Big|\boldsymbol{y}_{\mathrm{old}},\boldsymbol{\Theta}_i\right\}\spc.
\end{align}
This is, in fact, the usual model averaging estimator (see e.g. \citealt[Chapter~9]{Ghosh}).

\noindent If the set $\boldsymbol{X}_{\mathrm{new}}$ is known in advance, (\ref{eq:pred_conditional}) can be calculated on the fly (within each Metropolis iteration), and significant calculation time can be saved, depending on the acceptance rate of our sampling - provided that no new predictions are made whenever a new vector is rejected.

\noindent It is also worthwhile to consider ``denoising'' through elimination of highly unlikely models from (\ref{eq:pred_averaged}), by setting exceptionally small posterior probabilities (according to a predetermined threshold) to zero.

\subsection{Prediction based on Variable Selection}
\noindent Another obvious option is to regard the Bayesian mechanism of Section \ref{sec:Bayesian} as a mere selection procedure to identify influential (either linear or nonlinear) variables. In that case, inert variables in either the regression or the correlation part, will be set to $0$ or $1$, respectively, and the prediction at location $\bx$ will be
\begin{align}\label{eq:krig_nugget}
\widehat{y}\ofx = \bx^{\mathsf{T}}\widehat{\boldsymbol{\beta}} + \boldsymbol{r}_{\widehat{\boldsymbol{\rho}}}^{\mathsf{T}}\ofx\left[\boldsymbol{R}_{\widehat{\boldsymbol{\rho}}}+\widehat{\lambda}\boldsymbol{I}\right]^{-1}\left(\by-\boldsymbol{X}\widehat{\boldsymbol{\beta}}\right)\spc,
\end{align}
where $\left(\widehat{\boldsymbol{\beta}},\widehat{\boldsymbol{\rho}},\widehat{\lambda}\right)$ are the maximum likelihood estimators of $\left(\boldsymbol{\beta},\boldsymbol{\rho},\lambda\right)$, obtained by solving iteratively
$$
\begin{cases}
\widehat{\boldsymbol{\beta}}\left(\widehat{\boldsymbol{\rho}},\widehat{\lambda}\right) = \left(\boldsymbol{X}^{\mathsf{T}}\left[\boldsymbol{R}_{\hat{\boldsymbol{\rho}}}+\widehat{\lambda}\boldsymbol{I}\right]^{-1}\boldsymbol{X}\right)^{-1}\left[\boldsymbol{R}_{\hat{\boldsymbol{\rho}}}+\widehat{\lambda}\boldsymbol{I}\right]^{-1}\by \spc,\\[1em]
\widehat{\sigma}^2\left(\widehat{\boldsymbol{\beta}}, \widehat{\boldsymbol{\rho}},\widehat{\lambda}\right)=n^{-1}\left(\by-\boldsymbol{X}\widehat{\boldsymbol{\beta}}\right)^{\mathsf{T}}\left[\boldsymbol{R}_{\hat{\boldsymbol{\rho}}}+\widehat{\lambda}\boldsymbol{I}\right]^{-1}\left(\by-\boldsymbol{X}\widehat{\boldsymbol{\beta}}\right)\spc,\\[1em]
\left(\widehat{\boldsymbol{\rho}},\widehat{\lambda}\right)=\displaystyle\argmin_{\boldsymbol{\rho},\lambda}\left\{n\log\widehat{\sigma}^2\left(\widehat{\boldsymbol{\beta}}\left(\boldsymbol{\rho},\lambda\right), \boldsymbol{\rho},\lambda\right) + \log\left|\boldsymbol{R}_{\boldsymbol{\rho}}+\lambda\boldsymbol{I}\right|\right\}\spc.
\end{cases}
$$
Details of the optimization routine are given in \cite{DiceKriging}.\\

\noindent Denoting $\mathcal{M} = \left(\bgamma^r,\bgamma^c\right)$, our model of choice could then be the highest posterior probability model, i.e. $$\mathcal{M}^* = \displaystyle\argmax_{\mathcal{M}}p\left(\mathcal{M}\big|\by\right)$$ (to which we will refer in the text as the \textit{maximum a posetriori model}, or simply - the \textit{MAP model}), or, alternatively, we may set a threshold $q$ on the \textit{posterior inclusion probability}
$$
p_i = \sum_{\ell:\gamma_i=1}p\left(\mathcal{M}_{\ell}\big|\by\right)\spc,
$$
(where $\gamma_i$ may be either $\gamma_i^r$ or $\gamma_i^c$), and select the variable $x_i$ to either part, only if $p_i\geq q$. The choice of $q=\frac{1}{2}$ leads to what is famously known as the `median model', which - in the simpler case of linear regression with uncorrelated noise - has been shown by \cite{barbieri} to be the best predictive model, under very strict conditions on both the prior distributions and the covariates where predictions are to be made. In Section \ref{sec:bayes_var_simulation} we set the threshold on the posterior inclusion probability at a relatively high value of $0.8$.  A more educated choice of that value is proposed in Subsection \ref{subsec:Qian}, based on v-fold cross-validation.

\section{A Simulation Study}\label{sec:bayes_var_simulation}
\begin{figure}[ht]
\centering
 \includegraphics[width=1\linewidth]{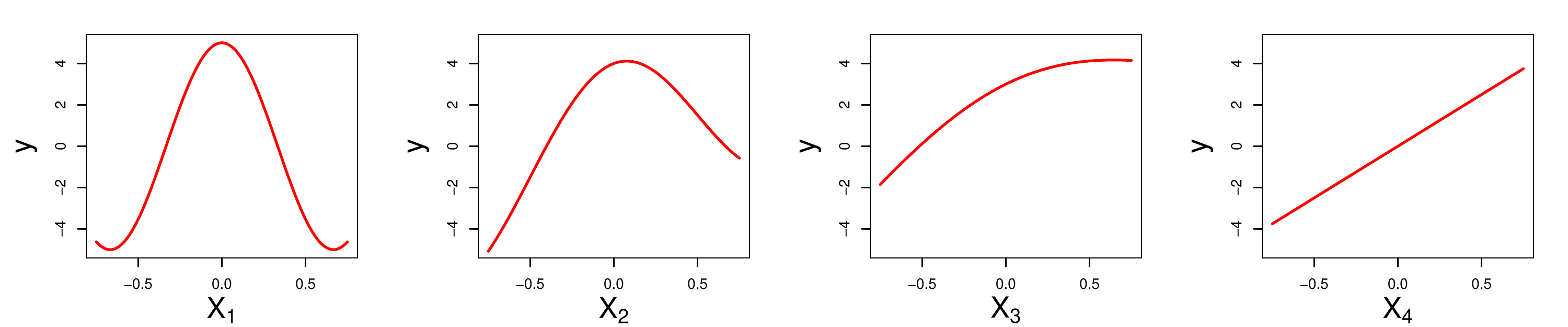}
  \caption{Plotting function (\ref{eq:bayes_var_select_sim}) against each of $X_1$ - $X_4$.}
  \label{fig:Simulation_Univariate_Comps}
\end{figure}

We now wish to test the proposed methods on simulated data. For that purpose, we used a $35$-run maximin Latin hypercube design (see \citealt{jones+johnson}) in $[-0.75,0.75]^5$ to sample noisy data from the allegedly $5$ dimensional function
\begin{align}\label{eq:bayes_var_select_sim}
y\ofx = 3X_2 + 4X_3 + 5X_4 + 5\cos\dfrac{3\pi X_1}{2} + 4\cos\dfrac{2\pi X_2}{2} + 3\cos\dfrac{\pi X_3}{2} + \varepsilon\ofx\spc,
\end{align}
where $\varepsilon\ofx\stackrel{\text{i.i.d.}}{\sim}\mathcal{N}\left(0,0.1^2\right)$ . Note that while (\ref{eq:bayes_var_select_sim}) consists of both linear and non-linear parts, the variables $X_1$ and $X_4$ are absent from the linear and non-linear parts, respectively, and $X_5$ is inert. Figure \ref{fig:Simulation_Univariate_Comps} shows how the different (additive) components of function (\ref{eq:bayes_var_select_sim}) vary as functions of their respective variables.\\

\noindent We assigned prior distributions as follows:
\begin{align*}
&\beta_j\big|\gamma_j^r=1\sim\mathcal{N}\left(0,5^2\right)\spc,\spc j=1,\ldots,5\spc;\spc\omega_r,\omega_c\sim\mathcal{U}(0,1)\spc;
\beta_0\sim\mathcal{N}\left(0,10^2\right)\spc;\\[0.5em]&\sigma^2_{_Z}\sim\text{Inv-}\Gamma(3,2)\spc\text{and}\spc\lambda\sim\text{Inv-}\Gamma(3,0.2)\spc,
\end{align*}
and calculated the root mean square prediction error (RMSPE) on an independent $100$-run maximin LHD in the same region.
A size $250,000$ sample was drawn from the posterior distribution, as described in Section \ref{sec:MCMC}, after the first $5000$ were discarded as burn-in.

\begin{figure}[ht]
\centering
\subfloat{\label{Beta_Simulation}\includegraphics[width=.5\linewidth]{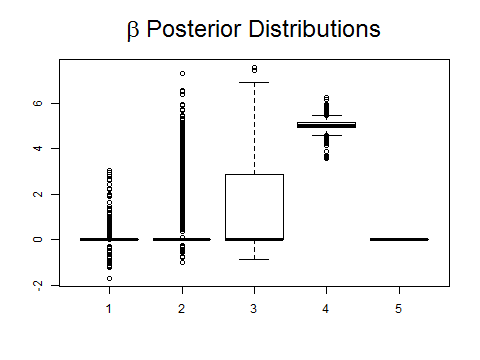}}
  \subfloat{\label{Rho_Simulation}\includegraphics[width=.5\linewidth]{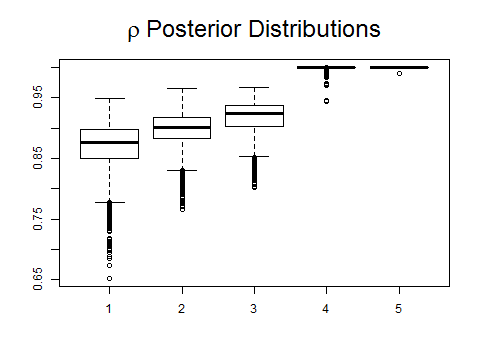}}
    \caption{Box plots of the posterior distributions of the regression and correlation parameters, for the simulated data.}
  \label{fig:Posterior_Boxplots_Simulation}
\end{figure}

\begin{figure}[ht]
 \includegraphics[width=1\linewidth]{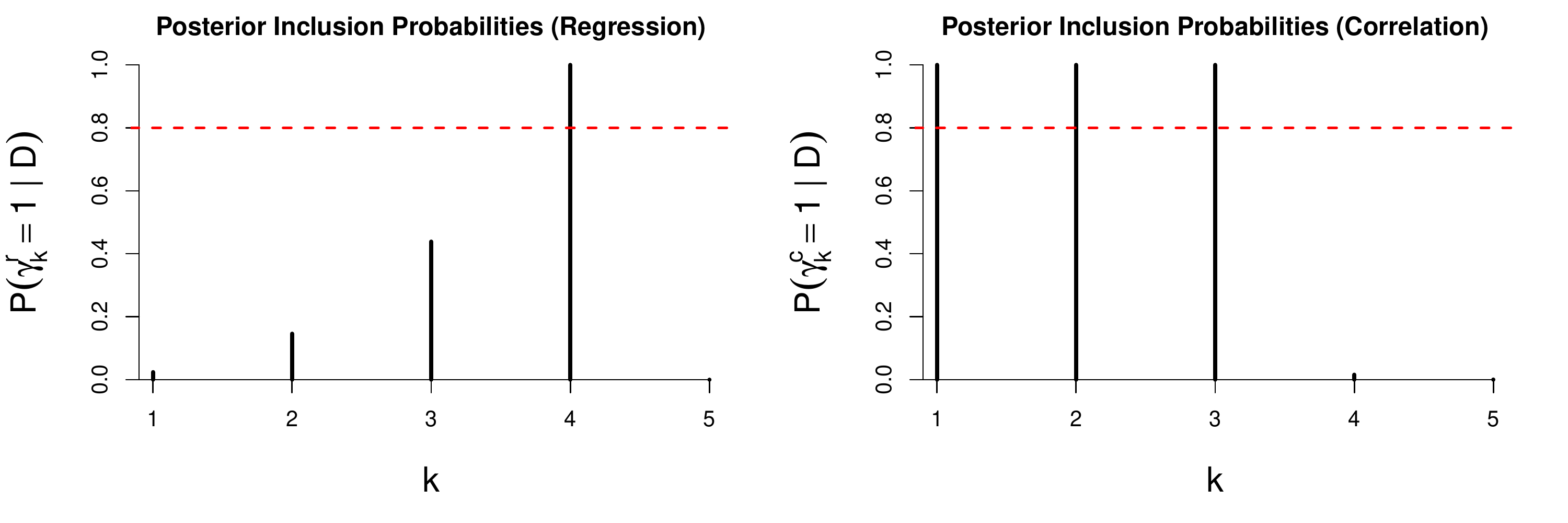}
  \caption{The posterior inclusion probabilities for the simulated data. The dashed line marks $80\%$ posterior inclusion frequency.}
  \label{fig:Simulation_Inclusion_Probs}
\end{figure}

\noindent Figure \ref{fig:Posterior_Boxplots_Simulation} shows the posterior distributions of both the regression coefficients and the correlation parameters, while Figure \ref{fig:Simulation_Inclusion_Probs} shows their respective marginal posterior inclusion probabilities. For now, we set the inclusion limit at $0.8$, leaving us with a single regression term ($X_4$) in the trend part of the model, and three covariates in the nonlinear component ($X_1$, $X_2$ and $X_3$). This is also consistent with Figure \ref{fig:Posterior_Boxplots_Simulation}. Apparently, the Gaussian process term effectively reflects the linear effects of $X_2$ and $X_3$ as well as their nonlinear effects, which explains why they were dropped from the regression component. As expected, Figure \ref{fig:Posterior_Boxplots_Simulation} clearly identifies $X_1$ as inactive in the trend part and $X_4$ in the non-linear component, while $X_5$ is deemed inert beyond any doubt.\\

\begin{figure}
  \centering
 \includegraphics[width=.45\linewidth]{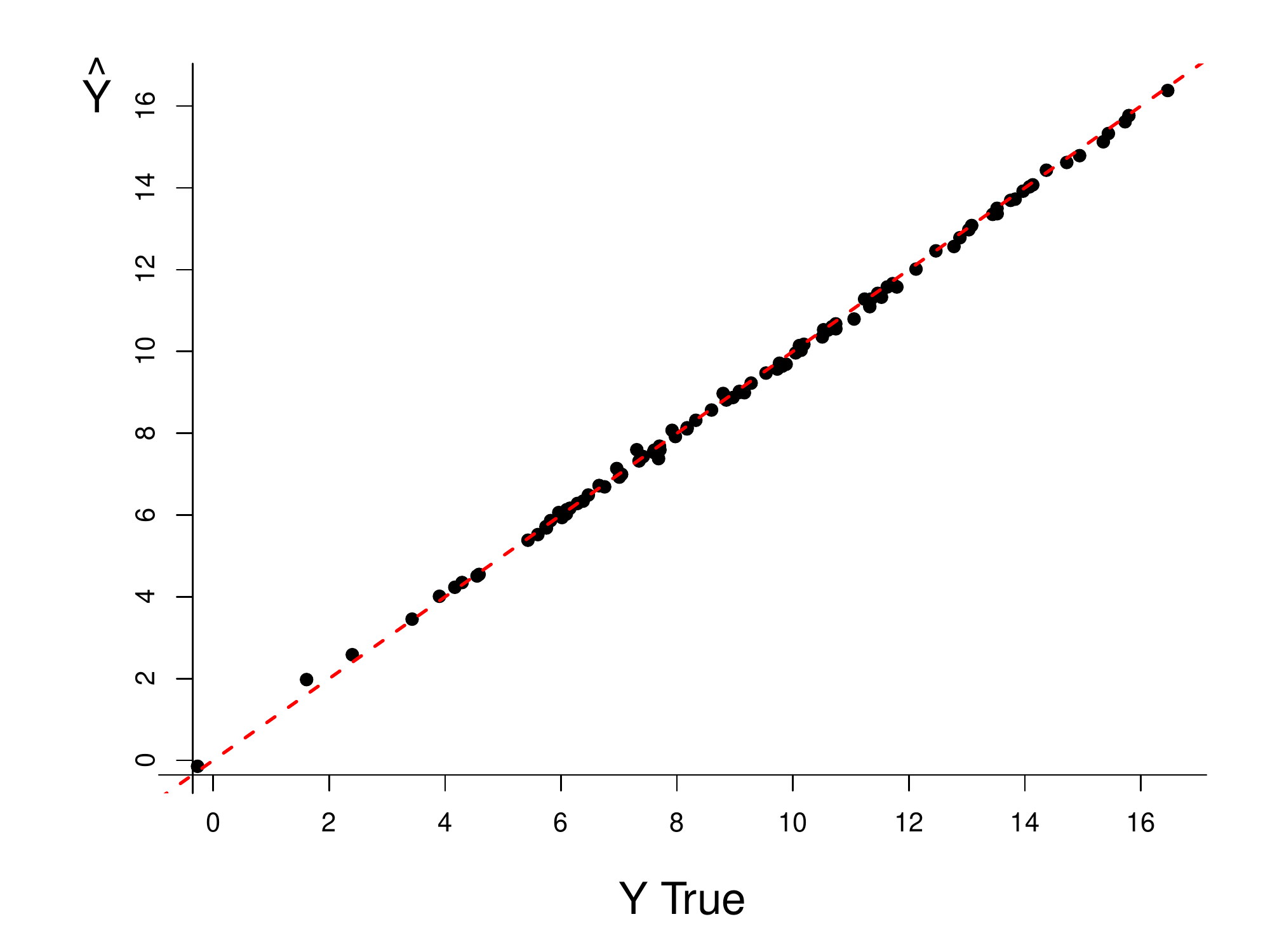}
  \caption{Predictions vs. true values for the simulated example.}
  \label{fig:Simulation_True_vs_Fitted}
\end{figure}

\noindent Table \ref{tab:Sim_Comp_Bayes_Var_Select} contains a comparison of the empirical RMSPE of several fitted models when tested on the $100$-run validation set.  The models include ordinary kriging (OK, no linear trend at all), universal kriging (UK, a linear trend which includes all of the covariates), model averaging based on (\ref{eq:pred_averaged}), variable selection based on the posterior inclusion probability ($0.8$ at least) and the empirical highest posterior probability (MAP) model.

\begin{table}[ht!]
\caption{Comparison of several prediction methodologies for the simulated data.}
\centering
{\footnotesize
\begin{tabular}{|c|c|}
\hline
{\bf\small Method} &{\bf\small RMSPE}\\
\hline
OK &$0.168$\\
\hline
UK &$0.130$\\
\hline
Averaging &$0.247$\\
\hline
Posterior Inclusion &$0.115$\\
\hline
MAP &$0.124$\\
\hline
\end{tabular}
}
\label{tab:Sim_Comp_Bayes_Var_Select}
\end{table}

\noindent Interestingly, the MAP model in this example was the one
containing $X_2$, $X_3$ and $X_4$ in its linear part, along with
$X_1$, $X_2$ and $X_3$ in its non-linear part, compatible with the
mechanism generating the data. However, it was outperformed by the
model selected according to the posterior inclusion probabilities.
It is our recommendation to resist the temptation to blindly opt for
the empirical MAP model, when it is unclear whether or not it is the
true highest posterior model - in particular when the posterior
distribution is multimodal or flat near its mode. In this example,
not for the only time in this paper, model averaging proved to be a
poor strategy. One possible reason could be poor estimation of the
posterior distribution, resulting in excessive weight on low
priority models. Figure \ref{fig:Simulation_True_vs_Fitted} 
demonstrates the performance on the hold-out set of the
model selected by the posterior inclusion probabilities.

\section{Applications}

\subsection{Heat Exchanger Emulator}\label{subsec:Qian}
\noindent We now wish to test our method on the computer simulation described
in \cite{Qian}. Here, the total rate of steady state heat transfer of a heat
exchanger for electronic cooling applications is modelled as a function
of four input factors: the mass flow rate of entry air, the temperature of entry air, the solid material thermal
conductivity and the temperature of the heat source. The response was evaluated at a $64$-run
experimental design, constituting an orthogonal array-based LHD, and produced by a finite difference
code. An additional $14$ run test set was used to validate the resulting emulator. Input variables were standardized into the $[0,1]$ region before analysis, to overcome the different scales. Using the prior distributions of Section \ref{sec:bayes_var_simulation}, a sample of size $155,000$ was drawn from the posterior distribution, with the first $5000$ serving as burn-in.\\

\noindent Previously, only interpolators have been attempted, as the simulation itself is deterministic. \cite{Qian} reported a RMSPE of $2.59$ using a universal Kriging model with a linear trend, compared to $5.15$ achieved by ordinary Kriging, while
\cite{CGP} reduced the RMSPE to $2.24$. Lately, \cite{CombinedGP} managed an RMSPE of $1.93$, using a convex combination of two independent Gaussian processes to model the spatial structure. Here we allowed for a non-zero nugget effect.  The different methods are compared  in Table \ref{tab:Qian_Comp_Bayes_Var_Select}.
One interesting feature is that $\bx_4$ has a strong linear effect but is slightly less dominant in the covariance.  The other factors are modeled primarily via the covariance. This slightly deviates from the work of \cite{Qian}, who used linear terms for $\bx_2$ and $\bx_4$, and modelled the correlation through all four factors.\\

\begin{figure}[ht]
  \centering
 \includegraphics[width=1\linewidth]{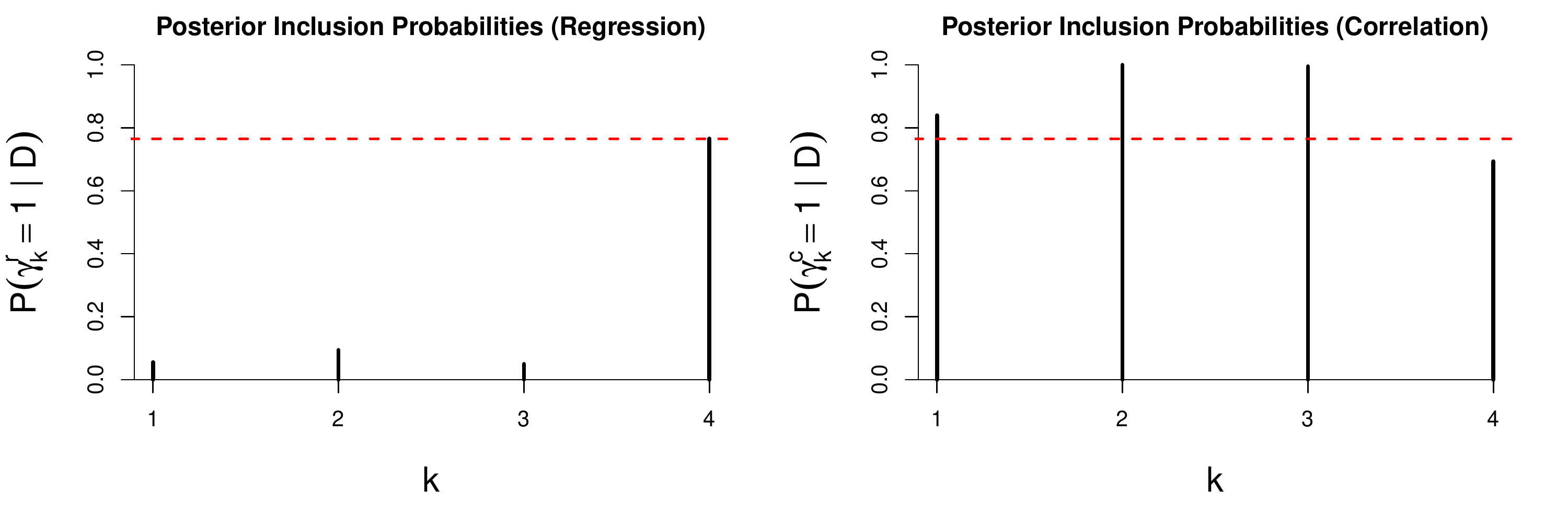}
  \caption{The posterior inclusion probabilities for the heat exchanger simulation. The dashed line marks a $77\%$ posterior inclusion level.}
  \label{fig:Qian_Inclusion_Probs}
\end{figure}

\begin{figure}[ht]
  \centering
 \includegraphics[width=.6\linewidth]{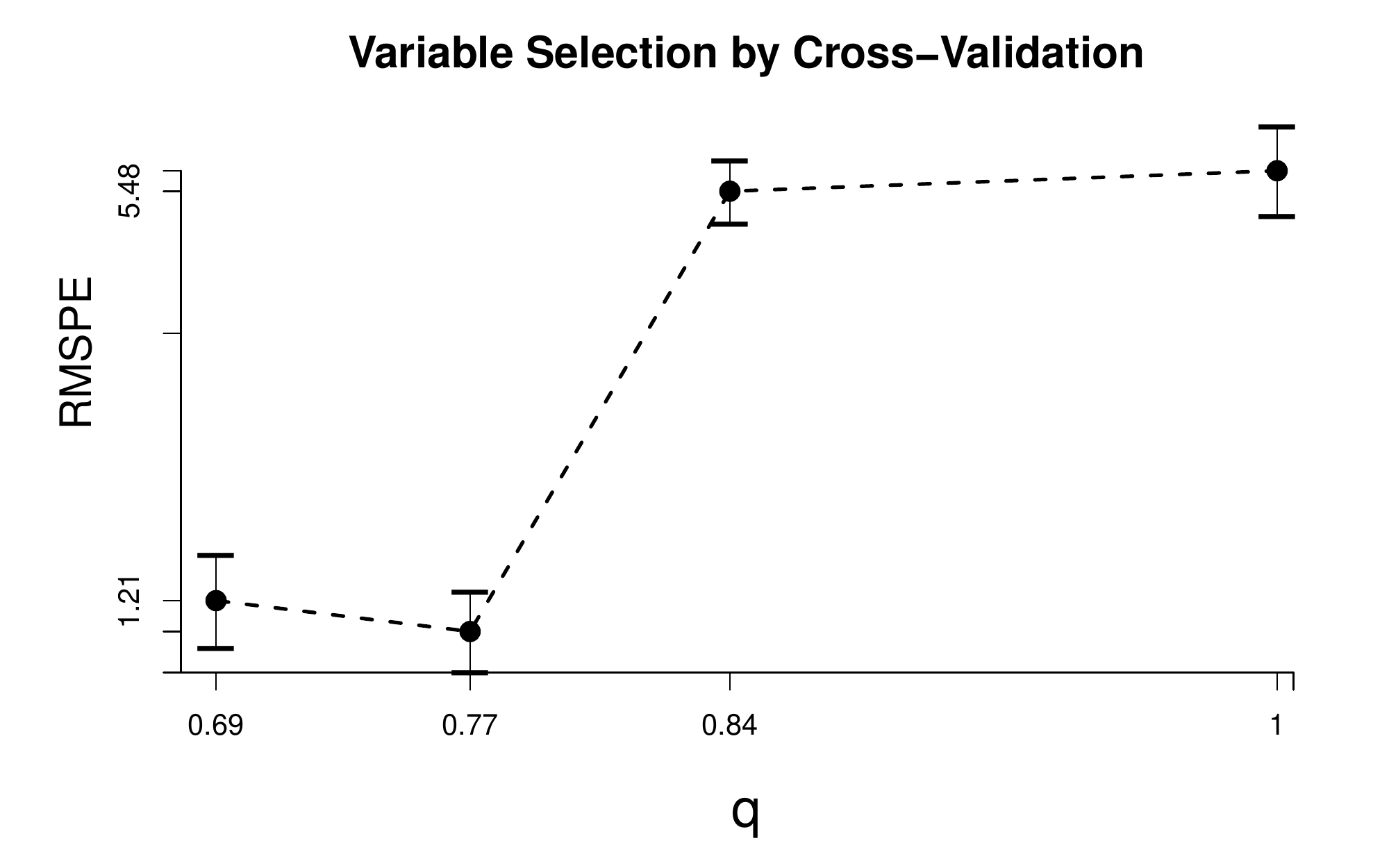}
  \caption{$8$-fold cross-validation results for the Heat Exchanger simulation data, including 1-SE marks.}
  \label{fig:Qian_CV}
\end{figure}

\noindent The dashed, horizontal line in Figure \ref{fig:Qian_Inclusion_Probs} marks the cutoff used in this example to omit terms from the model. The cutoff - $77\%$ - was obtained via the following procedure:  only terms with at least a $30\%$ marginal inclusion probability were considered, and terms with $90\%$ marginal inclusion probability or more were automatically included. That left $4$ competing models, of which the eventual model was chosen via $8$-fold cross-validation (fitting the model, based on $56$ observations and testing on the remaining $8$ each time). Figure \ref{fig:Boston_CV} shows the results, along with one-standard error marks, potentially serving to prioritize sparse models, in a similar manner to CART models (see \citealt[chapter~3]{cart}). Generally, using this procedure leaves us with at most $2p$ models to compare (and, in practice, typically far less), instead of $2^{2p}$ in an exhaustive search. As evident from Figure \ref{fig:Qian_CV}, in this example, a model consisting of $4$ predictors ($X_1$ in the linear part and $X_2$, $X_3$ and $X_4$ in the stochastic part), was selected.\\

\begin{figure}[ht]
\subfloat{\label{Beta_Qian}\includegraphics[width=.5\linewidth]{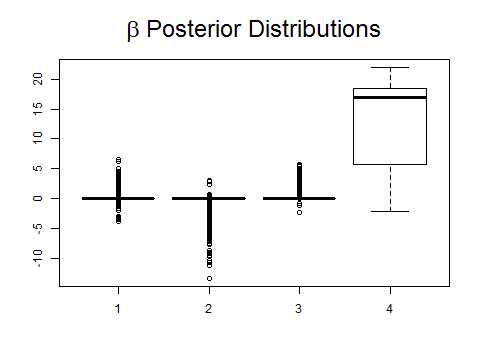}}
  \subfloat{\label{Rho_Qian}\includegraphics[width=.5\linewidth]{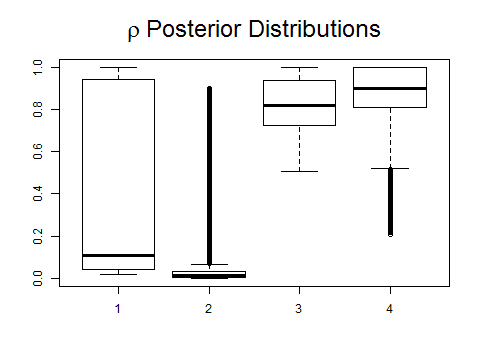}}
    \caption{Box plots of the posterior distributions of the regression and correlation parameters for the heat exchanger data.}
  \label{fig:Posterior_Boxplots_Qian}
\end{figure}

\begin{table}[ht!]
\caption{Comparison of several prediction methodologies for the heat exchanger example.}
\centering
{\footnotesize
\begin{tabular}{|c|c|}
\hline
{\bf\small Method} &{\bf\small RMSPE}\\
\hline
OK &$2.023$\\
\hline
UK &$2.670$\\
\hline
Averaging &$2.686$\\
\hline
Postrior Inclusion &$1.793$\\
\hline
MAP &$1.793$\\
\hline
\end{tabular}
}
\label{tab:Qian_Comp_Bayes_Var_Select}
\end{table}

\subsection{The Boston Housing Data}\label{subsec:boston}
\begin{figure}[ht]
\centering
\subfloat{\label{Beta_Boston}\includegraphics[width=.5\linewidth]{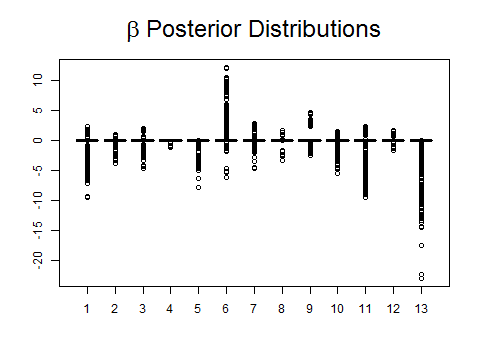}}
  \subfloat{\label{Rho_Boston}\includegraphics[width=.5\linewidth]{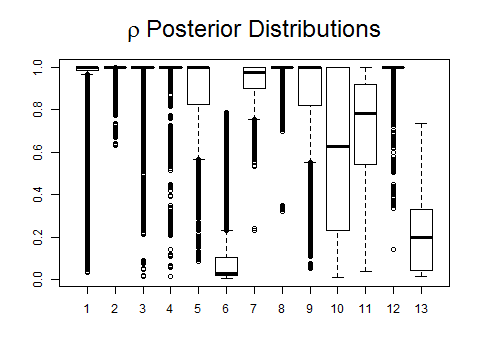}}
    \caption{Box plots of the posterior distributions of the regression and correlation parameters, for the Boston Housing data.}
  \label{fig:Posterior_Boxplots_Boston}
\end{figure}

\noindent To test our method on a larger-scale problem, we now
analyze the Boston Housing data set, which also appeared in
\cite{Vannucci}, and - most famously - in \cite{Breiman}, where a
full description of the $13$ predictors - related to the median
value of owner-occupied homes in census tracts of the Boston
metropolitan area - can be found. The data set contains $506$
observations, of which we used $130$ as a training set and the other
$376$ for validation. Again, we transformed the predictors so that
they are all in the $[0,1]$ interval. We produced a sample from the
posterior distribution of size $250,000$ (with an additional burn-in
sample of size $15,000$), and the same set of priors as in Section
\ref{sec:bayes_var_simulation}. The resulting marginal posterior
distributions are shown in Figure
\ref{fig:Posterior_Boxplots_Boston}, where it is evident that none
of the covariates makes a pure, linear contribution to the median
property value, and that the $6$th (average number of rooms per
dwelling), $10$th (full-value property-tax rate per $\$10,000$),
$11$th (pupil-teacher ratio by town) and $13$th (\% lower status of
the population) covariates stand out in the non-linear part of the
model.

\begin{figure}[ht]
  \centering
 \includegraphics[width=1\linewidth]{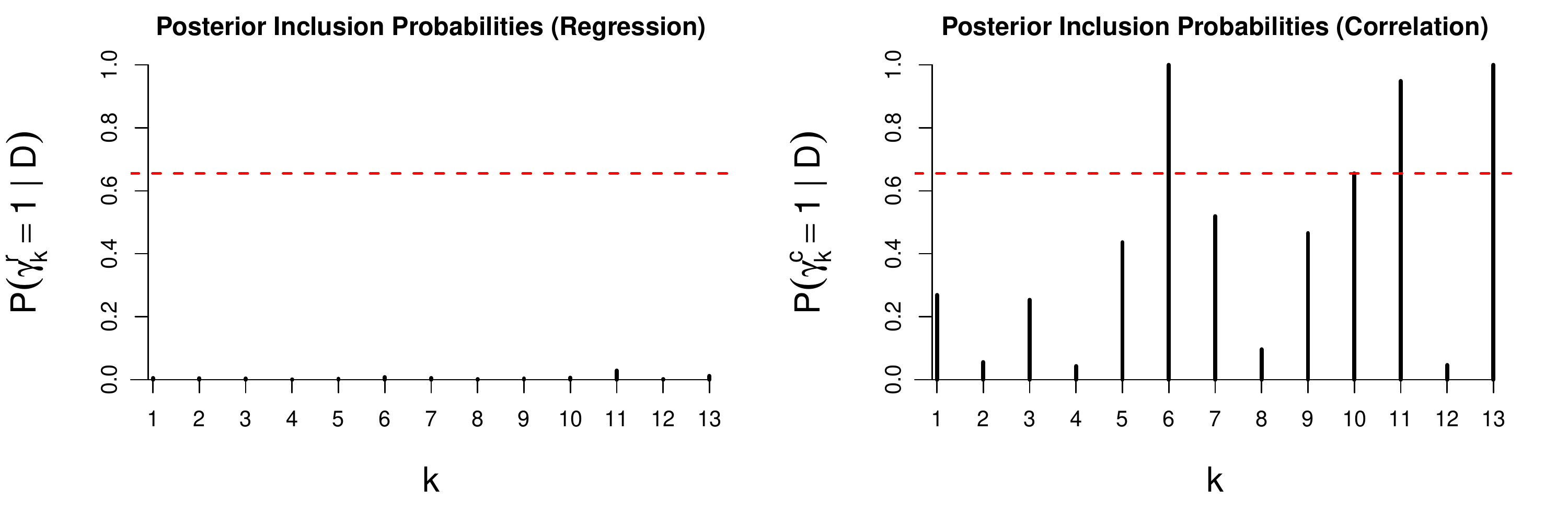}
  \caption{The posterior inclusion probabilities for the Boston Housing data set. The dashed line marks a $76\%$ posterior inclusion frequency.}
  \label{fig:Boston_Inclusion_Probs}
\end{figure}

\begin{figure}[ht]
  \centering
 \includegraphics[width=.6\linewidth]{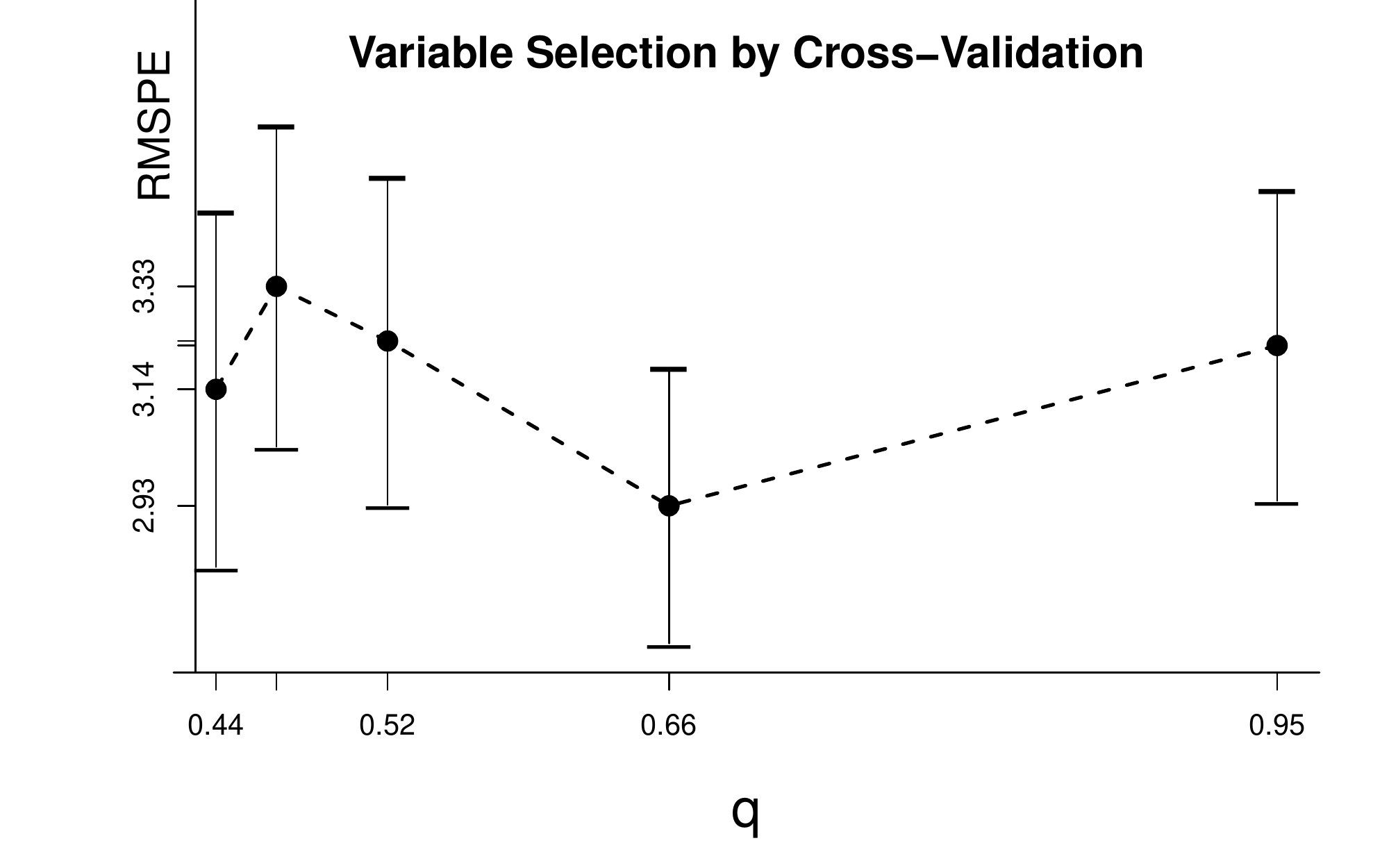}
  \caption{10-fold cross-validation results for the Boston Housing data, including 1-SE marks.}
  \label{fig:Boston_CV}
\end{figure}

\noindent Figure \ref{fig:Boston_Inclusion_Probs}, where the marginal posterior inclusion probabilities are shown, is very much in line with Figure \ref{fig:Posterior_Boxplots_Boston}. As previously in \ref{subsec:Qian}, the cutoff - $66\%$ - was determined through a $10$-fold cross-validation procedure, shown in Figure \ref{fig:Boston_CV}. In this example, the most parsimonious model, consisting of only $4$ predictors, was singled out as the obvious choice.\\

\begin{figure}
  \centering
 \includegraphics[width=1\columnwidth]{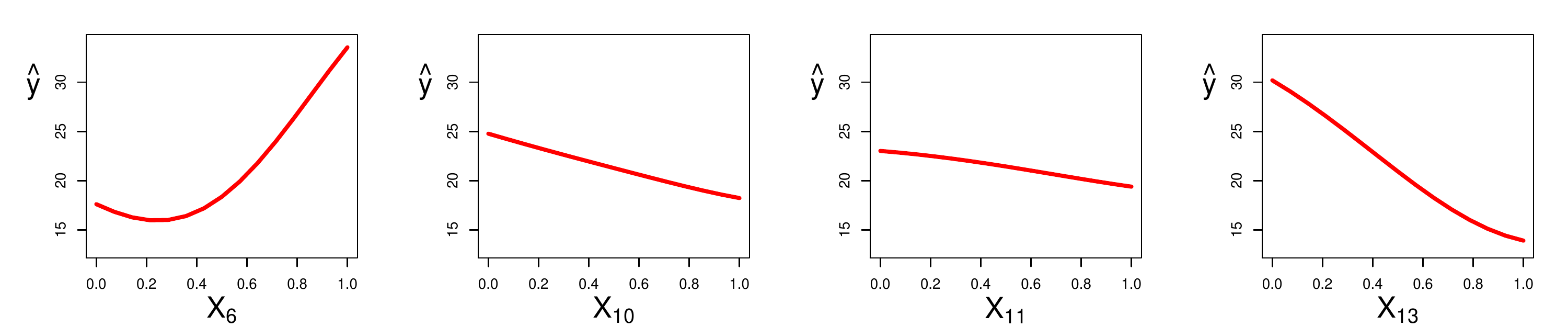}
  \caption{The predicted value $\widehat{y}\ofx$ for the Boston Housing data, plotted against $X_6$, $X_{10}$, $X_{11}$ and $X_{13}$, averaged over all other variables in each plot.}
  \label{fig:Boston_Univariate_Predictions}
\end{figure}

\noindent In Figure \ref{fig:Boston_Univariate_Predictions}, the reader can see how the emulator values $\widehat{y}\ofx$ vary for each of the covariates. We evaluated (\ref{eq:krig_nugget}) at numerous sites on a regular grid, and the plotted values in each graph were obtained by averaging on the other three variables. It should come as no surprise to the reader, based on Figure \ref{fig:Posterior_Boxplots_Boston}, that $X_6$ and $X_{13}$ are the variables accounting for the dominant nonlinear effects, while the output seems to be far less sensitive to changes in the values of $X_{10}$ and $X_{11}$, consequence of their $\widehat{\rho}$ estimates of $0.392$ and $0.668$, respectively. Thorough discussion of sensitivity analysis can be found in \cite{saltelli}. 
While the reader may be misled by Figure \ref{fig:Boston_Univariate_Predictions} to think that $X_{10}$ and $X_{11}$ are linear 
effects in this example, we feel the need to emphasize that these plots were obtained by averaging, and do not capture interactions. That also explains the flat (or even slightly decreasing) average behavior of $\widehat{y}\ofx$ for the lower values of $X_6$ (average number of rooms per
dwelling).

\noindent It is interesting to note that while the ordinary kriging
and universal kriging models achieved empirical RMSPEs of $4.02$ and
$4.01$, respectively, on the hold-out set, the selected model, based
on $4$ predictors, achieved a RMSPE of $3.98$. Although prediction
is not always the main purpose of variable selection procedures of
this sort, it is still encouraging to find out that dimensionality
reduction of this magnitude does not necessarily come at the expense
of predictive accuracy. For further reference, we used the Treed
Gaussian Process (TGP) model, a nonstationary model, developed and
implemented by \cite{TGP}, which combines partitioning of the
predictor space with local, stationary GP models. In this example,
the TGP model achieved a RMSPE of $4.16$.

\section{Discussion}
\noindent In this work, we explored variable selection in Bayesian semiparametric regression models.
Although the primary goal of these procedures is often of qualitative or explanatory nature, we keep one eye open
on the accuracy of our predictions. With that in mind, our variable selection procedure is driven by prediction
throughout, as evident from the cross-validation procedure described in Subsections \ref{subsec:Qian} and \ref{subsec:boston}, and unlike the frequentist-like selection procedure of \cite{Bingham}, where identification of the active inputs was the sole purpose.\\

\noindent Ideally, we would like to take advantage of
the Bayesian framework to provide accurate predictions, using (\ref{eq:pred_averaged}). However, using model averaging to a good effect might require
an educated choice of prior distributions, along with possibly vast samples from the posterior disributions. One very interesting open question is whether a set of sparsity prior distributions exists, such that under some aggregation methods, the minimax convergence rate of the mean squared prediction error can be achieved, as in the exponential weighting scheme of \cite{Tsybakov} for linear regression models.\\

\noindent Naturally, one may consider replacing model (\ref{eq:model}) with universal kriging models of the form
$y\ofx = \beta_0 + \boldsymbol{f}\ofx^{\mathsf{T}}\boldsymbol{\beta} + Z\ofx + \varepsilon\ofx$,
where $\boldsymbol{f}\ofx = \left[f_1\ofx,\ldots,f_r\ofx\right]^{\mathsf{T}}$ for some dictionary of regression functions. In that case, extra caution needs to be taken, as choice of regression functions already in the reproducing kernel Hilbert space associated with the covariance kernel $R\left(\cdot,\cdot\right)$ could cause a multicollinearity-like effect (see \citealt{Dizza}), resulting in the omission of crucial factors from our model.\\

\noindent One question left unanswered in this paper, and which may
be the basis for future research, is whether or not one should
include some linear terms for extrapolation purposes, even when the
selection process indicates otherwise. Based on the artificial
example of Section \ref{sec:bayes_var_simulation}, it looks very
likely that the model will drop most of the regression functions and
rely on the covariance kernel. It would therefore be interesting to
see how devastating an effect that would have on predictions outside
the convex hull of the training data. Another possibility is to begin 
by ``orthogonalizing'' the covariance kernel to linear terms, by subtracting 
out the initial components of its Mercer expansion. Using Mercer's expansion in 
computer experiments has been covered by \cite{OptimalGP} and \cite{Muller2}, among others.

\section*{Supplementary Material}
\noindent \textbf{\textsf{R} code and data files:} \textsf{R} code for all sections and data files can be downloaded by the readers as a .zip file.


\bibliographystyle{elsarticle-harv}
\bibliography{References}   

\end{document}